\documentclass[reprint,superscriptaddress,aip,jcp,]{revtex4-2}
\usepackage{graphicx}
\usepackage{placeins}
\usepackage{color}
\usepackage{latexsym}
\usepackage{amsmath}
\usepackage{amsfonts}
\usepackage{amssymb}
\usepackage{bm}
\usepackage{graphicx}
\usepackage{mathtools}
\usepackage{amsbsy}
\usepackage{physics}
\usepackage{enumitem}
\usepackage{lipsum}
\usepackage{rotating}

\expandafter\ifx\csname package@font\endcsname\relax\else
 \expandafter\expandafter
 \expandafter\usepackage
 \expandafter\expandafter
 \expandafter{\csname package@font\endcsname}%
\fi
\usepackage[normalem]{ulem}

\graphicspath{{./figures/},{figures}}

\newcommand{\lla}{\left\langle}
\newcommand{\rra}{\right\rangle}

\let\Gamma\varGamma
\let\Delta\varDelta
\let\Theta\varTheta
\let\Lambda\varLambda
\let\Pi\varPi
\let\Sigma\varSigma
\let\Upsilon\varUpsilon
\let\Phi\varPhi
\let\Psi\varPsi
\let\Omega\varOmega
\let\phi\varphi

\newcommand{\REV}[1]{\textcolor{black}{#1}}

\usepackage{hyperref}

\begin{document}
\title{Analytical Analysis of the Conformational and Rheological Properties of Flexible Active Polar Linear Polymers under Shear Flow}

\author{Arindam Panda}%
\email{arindam19@iiserb.ac.in}
\affiliation{Department of Physics, Indian Institute of Science Education and Research, \\ Bhopal 462 066, Madhya Pradesh, India}
\author{Sunil P Singh}
 \email{spsingh@iiserb.ac.in}
\affiliation{Department of Physics, Indian Institute of Science Education and Research, \\ Bhopal 462 066, Madhya Pradesh, India}
\author{Roland G. Winkler}%
\email{rg\_winkler@gmx.de}
\affiliation{Theoretical Physics of Living Matter, Institute for Advanced Simulation, Forschungszentrum J{\"u}lich, 52425 J{\"u}lich, Germany}

\date{\today}

\begin{abstract}
The conformational and rheological properties of active polar linear polymers (APLPs) under linear shear flow are studied
analytically. We describe a discrete APLP as an inextensible flexible Gaussian bead-spring chain supplemented by active forces along the bonds. The
linear, non-Hermitian equations of motion are solved by an eigenfunction expansion in terms of a biorthogonal basis set. The model reveals an intimate coupling between activity and shear flow, which implies activity-enhanced polymer conformational and rheological properties. 
Compared to a passive polymer, we find a significantly enhanced shrinkage transverse to the flow direction with increasing shear rate, with a power-law exponent $-4/3$, compared to the passive values of $-2/3$.  This conformational change is tightly linked with a strongly amplified shear-thinning behavior, where the shear viscosity exhibits the same power law. The characteristic shear rate for the onset of these effects is determined by the activity. In the asymptotic limit of large activities, the shear-induced features become independent of activity and equal to those of passive polymers.
\end{abstract}

\maketitle

\section{Introduction}

Nature provides a wide spectrum of filament- and polymer-like structures, which either move autonomously or are propelled by the coupling with their environment.\cite{wink:20} Many motile organisms self-organize into chain-like assemblies \cite{sela:11,sohn:11,yama:19,love:19,wink:20} leading to the formation of biofilms, reticulate patterns, and scaffold-like structures at macroscopic scales.\cite{sugi:19} Cellular functions crucially depend on the out-of-equilibrium and active motion of biopolymers. A paramount example are polar filaments such as actin and microtubules, which are driven by molecular motors.\cite{nedl:97,howa:01,baus:06,rama:10,rama:10,sanc:12,marc:13,pros:15,ravi:17,wink:20} Actin filaments and networks composed of them reveal complex activity-induced mechanical responses, such as large-scale buckling, reduced persistence length or flexibility, and a significantly enhanced network elastic module,\cite {live:01,scha:10,butt2010myosin,sumi:12,xu2024self} to name a few features. Within the nucleus, replication of DNA by DNA polymerase or transcription of RNA by RNA polymerase generates nonthermal fluctuations.\cite{albe:22,guth:99,meji:15,dipi:18} Moreover, active processes are suggested to be essential for the arrangement of the eukaryotic genome, controlling their
dynamical properties, and to be essential for the cell functions.~\cite{dipi:18,jave:13,zido:13,lieb:09,gana:14,smre:17,wink:20}

The active filaments and polymers generate a flow in their embedding fluid environment, which can play a major role in their transport properties.\cite{sain:18.1} Even more, external flows can create forces and torques that affect the polymer conformations, dynamics, and transport.\cite{whee:19} An example is semiflexible nematodes, which reveal an intriguing shear-thinning behavior and a significantly faster viscosity drop with increasing shear rate than passive polymers.\cite{back:13,malv:19} Similarly, the shear-thinning behavior of compact, highly entangled blobs of living {\em T. tubifex} worms differs quantitatively from that of passive polymers.\cite{debl:20.1,debl:20}   

Studies on passive flexible and semiflexible polymers under steady shear-flow uncover fascinating structural and dynamical nonequilibrium features.  \cite{lars:99,larson2005rheology,wink:06.1,schr:05.1,doyl:97,smit:99,liu:18.1} Fluorescence light scattering experiments on DNA and actin filaments reveal a cyclic polymer stretching and recoiling/folding---denoted as tumbling---, due to their major conformational fluctuations over time, rather than maintaining a stationary stretched state.\cite{schr:05.1,hur:01,larson2005rheology} Molecular simulations\cite{aust:99,huan:10,huan:11,dala:12,delg:06,sing:13,lamu:12,lang:14,liu:18.1,lamu:21,saju2024dry} and theoretical analysis\cite{cher:05,wink:06.1,wink:10} have shown that this cyclic motion is a consequence of a continuous stretching and compression/folding of a polymer. Moreover, the polymer conformational changes are tightly linked to their rheological behavior, specifically stretching along the flow direction and shrinkage in the transverse direction, which leads to a pronounced shear thinning.~\cite{aust:99,huan:10,wink:10,sing:13}

The coupling of the two nonequilibrium phenomena---internal polymer activity and external flow field---leads to additional effects.
In the strive to elucidate the microscopic mechanism of such an interplay, and to arrive at a rational understanding of the polymers' rheological behaviors, two main models are applied. On the one hand, polymers (linear or ring) are composed of active Brownian particles (APBs).\cite{tenh:11,bial:12,hows:07,gomp:09,bech:16,samanta2016chain} They are denoted as active Brownian polymers (ABPOs)\cite{kais:14,eise:16,wink:20,anan:20} or active Brownian ring polymers (ABRPs).\cite{mous:19} On the other hand, activity is applied tangentially~\cite{bian:18,janz:25} or along the bonds of a polymer,\cite{isel:15,phil:22.1,phil:22,faze:23,kuma:24,teje:24,wink:24,lamu:24,pand:25,wink:25,karan2024inertia,panda2025folding} the latter are denotes as active polar linear polymers (APLPs) and active polar ring polymers (APRPs).\cite{phil:22,wink:24} Theoretical~\cite{mart:18.1,wink:24} and simulation~\cite{pand:23,kuma:24,pand:25} studies have addressed the influence of shear flow, active forces, and their coupling onto the polymer properties for both kinds of models. Qualitatively, comparable features to those of passive polymers are observed, however, activity implies major quantitative differences, specifically for the shear-dependent shrinkage transverse to the flow direction, the shear viscosity, and the tumbling time. This applies to the polymer geometry---linear vs. ring---,\cite{wink:10,wink:24,kuma:24} as well as to the applied active force---active Brownian beads vs. tangential drive.\cite{mart:18.1,pand:25} With respect to the shear viscosity, activity leads to an enhanced shear thinning, and the power-law decreases of the viscosity with increasing shear rate, $\dot \gamma$, changes from $\dot \gamma^{-1/2}$ of a passive self-avoiding polymer\cite{huan:10} to $\dot \gamma^{-3/4}$ for an ABPO,\cite{pand:23} and to $\dot \gamma^{-4/3}$ for an APLP.\cite{pand:25}         

Until now, there has been a lack of theoretical knowledge about the shear-induced properties of APLPs. To fill this gap, we apply a minimal model of a tangentially driven polymer and analyze the interaction between shear flow and activity on its conformational and rheological properties.
The discrete polymer is modeled as a Gaussian flexible phantom bead-spring chain of finite extensibility in the absence of hydrodynamic interactions.~\cite{wink:94,wink:10,mart:18.1} Activity is applied along the bonds between successive beads, which is considered suitable for long polymer objects such as microtubules, actin filaments, slender bacteria, worms, or {\em Plasmodium} sporozoites. The polar interaction breaks the polymer end-to-end symmetry, with corresponding non-Hermitian, but linear Langevin equations, which are solved by an expansion into a biorthogonal basis set.\cite{phil:22.1}  

Our calculations reveal substantial deviations of an APLP from passive polymer behavior, providing detailed insight into the effects of activity and flow. Activity enhances the polymer deformation and the shear thinning behavior in an intermediate regime of shear rates, as observed in previous simulations.\cite{pand:25}  In the limit of strong shear, shear dominates over activity, and the active polymer behaves as a passive polymer. The theoretical analysis reveals a tight link between the shrinkage of the active polymer transverse to the flow direction and its shear viscosity, which both obey the approximate power law \REV{$\dot \gamma^{-4/3}$} in the activity-dependent, intermediate shear-rate regime. Moreover, our analysis provides scaling relations in terms of the shear rate and the activity for the crossover from weak shear to shear-induced deformations, as well as the crossover to passive polymer behavior at large shear rates. 

\begin{figure}[t]
\includegraphics[width=\columnwidth]{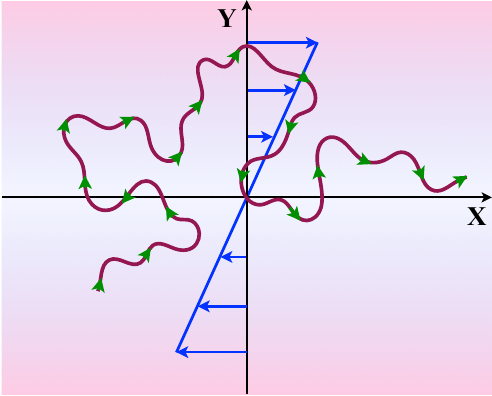}
	\caption{ A continuum representation of an active polymer under linear shear flow. The green arrows show the direction of the active force, while the blue arrows indicate the direction of the imposed shear flow. }
	\label{Fig:sketch}
\end{figure}

\section{Model}

\subsection{Equations of Motion}

The flexible active polar linear polymer (APLP) is  composed of $N+1$ beads with the time-dependent, $t$, positions $\bm r_i(t)$ ($i \in \{0,\ldots,N\}$) embedded in the three-dimensional space (Fig.~\ref{Fig:sketch}). Their overdamped equations of motion are given by   
\begin{align} \label{eq:eom_0}
    \zeta \frac{d}{dt} \bm r_0(t) = & \ \left(\frac{f_a}{2} + \frac{3k_BT\mu}{l^2}\right) \bm R_1(t)+ \zeta \mathrm{{\mathbf K}} \bm r_0(t) + \bm \varGamma_0 (t) , \\  \nonumber
    \zeta \frac{d}{dt} \bm r_i(t)  = & \ \frac{f_a}{2} \left( \bm R_{i+1}(t) + \bm R_{i}(t) \right) \\  \label{eq:eom_i}
    & \ \hspace*{-1.3cm} +  \frac{3k_BT\mu}{l^2} \left( \bm R_{i+1}(t) - \bm R_{i}(t) \right)  + \zeta \mathrm{{\mathbf K}} \bm r_i(t) + \bm \varGamma_i(t) , \\ \label{eq:eom_n}
    \zeta \frac{d}{dt} \bm r_N  = & \ \left(\frac{f_a}{2} - \frac{3k_BT
    \mu}{l^2} \right) \bm R_N(t)+ \zeta  \mathrm{{\mathbf K}}  \bm r_N(t) + \bm \varGamma_N(t) ,
\end{align}
with the friction coefficient $\zeta$, the Boltzmann factor $k_B$, and the temperature $T$. The linear bond forces, proportional to the strength $\mu$, are described within the Gaussian polymer model.\cite{wink:94} The stretching coefficient  $\mu$ (Lagrangian multiplier) itself accounts for the inextensibility of the polymer contour and is determined by the constraint
\begin{equation} \label{eq:const}
    \sum_{i=1}^N \lla \bm R_ i^2 \rra = N l^2  
\end{equation}
on the bond vectors, where $l$ is the bond length.\cite{wink:94,wink:24} The homogeneous active forces of magnitude $f_a$ act along the bond vectors $\bm R_i=\bm r_{i+1} - \bm r_i$ between successive beads. Shear is applied along the $x$-axis of the Cartesian reference frame, and the gradient direction is along the $y$-axis (Fig.~\ref{Fig:sketch}). Hence, only the component $K_{xy} = \dot \gamma$ of the shear rate tensor $\mathrm{{\mathbf K}}$ is different from zero. The stochastic forces $\bm \varGamma_i(t)$ describe the thermal fluctuations and are assumed
to be stationary, Markovian, and Gaussian with zero mean and the second moments
\begin{equation}
    \lla \bm \varGamma_i(t) \cdot \bm \varGamma_j (t') \rra = 6 \zeta k_BT \delta_{ij} \delta(t-t') .
\end{equation}

\subsection{Solution of the Equations of Motion}

The linear but non-Hermitian Langevin equations Eqs.~\eqref{eq:eom_0} -- \eqref{eq:eom_n} are solved by an eigenfunction expansion in terms of a biorthonormal basis set ${\bm b}_n=(b_n^0,b_n^1,\ldots,b_n^N)^T$ and ${\bm b}_n^{\dag}=(b_n^{0^\dag},b_n^{1\dag},\ldots,b_n^{N\dag})^T$ of the nonsymmetric matrix $\mathbf{M}$ of Appendix \ref{app:eigenvalues} and its adjunct, \cite{phil:22.1,pand:25} i.e.,
\begin{align} \label{eq:exp_pos}
\bm r_{i}(t) = & \ \sum_{n=0}^{N} \bm \chi_{n}(t) b_n^i, \\   \label{eq:gamma}
\bm \varGamma_{i} (t) = & \ \sum_{n=0}^{N} \Tilde{\bm \varGamma}_{n}(t)b_n^i,    
\end{align}
with the mode amplitudes $\bm \chi_{n}$ and $\Tilde{\bm \varGamma}_{n}$, respectively. The basis functions are  given by 
\begin{align} 
	&b_n^{i} = \sqrt{\frac{2}{N+1}} \frac{e^{-d i}}{\sqrt{(1-r^2) \sin^2 k_n + (r - \lambda_n/2)^2}} 
	\label{align:eigen_dis_b}
	\\		
	& \times \Big[ \sqrt{1-r^2} \sin k_n \cos(k_n i) + (r - \lambda_n/2) \sin(k_n i) \Big], \nonumber
	\\[0.5em]
	&b_n^{i^{\dagger}} = e^{2 d i} b_n^{i}, \hspace{1cm} n \in \mathbb{N}_0,
	\label{align:eigen_dis_b_dag}
	\\[0.5em] 
	&b_0^{i} = \sqrt{\frac{\sinh(d)}{e^{dN}\sinh(d(N+1))}}, 
	\label{align:eigen_dis_norm}
	\\[0.5em]
	&d = \ln (\sqrt{1+r}/\sqrt{1-r}), 
	\hspace{1.44cm}
	r <1 , \\
	&d = \ln (\sqrt{1+r}/\sqrt{r-1}) - \imath \frac{\pi}{2}, 
	\hspace{0.6cm}
	r >1,  
\end{align}	
with the wave numbers $k_n = n \pi / (N+1)$, the P\'eclet number 
\begin{equation} \label{eq:peclet}
    Pe = \frac{f_a l^2}{k_B T} ,
\end{equation}
and the abbreviation 
\begin{equation} \label{eq:abb_d}
    r = \frac{f_a l^2}{6 k_B T \mu} = \frac{Pe}{6 \mu} .    
\end{equation}
\REV{The conformations of an APLP in the absence of shear are independent of activity, i.e., $\mu=1$.\cite{phil:22.1} Hence, the ratio $r$ is simply a reduced P\'eclet number. Shear flow causes conformational changes, which implies $\mu \geq 1$, and $r$ captures the influence of activity and shear flow on the polymer properties. Specifically, in the limit $\mu \to \infty$, i.e., $r \to 0$, the eigenfunctions and eigenvalues turn into those of a passive polymer for any finite P\'eclet number.}  

By inserting the eigenfunction expansion in Eqs.~\eqref{eq:eom_0} -- \eqref{eq:eom_n}, we obtain the equations of motion for the mode amplitudes
\begin{equation} \label{eq:mode_eom}
    \zeta \frac{d }{dt} \chi_{\alpha n}(t) = -\xi_{n}\chi_{\alpha n}(t) + \Tilde\varGamma_{\alpha n}(t)+  \dot \gamma   \zeta\delta_{x \alpha }\chi_{y n}(t)  ,
\end{equation}
with $\alpha \in \{x,y,z\}$ and the eigenvalues $\xi_n$. The latter are given by 
\begin{equation} \label{eq:eigenvalues}
   \xi_n = \frac{3 k_BT \mu}{l^2} \lambda_n =   \frac{6 k_BT \mu}{l^2} \left( 1 - \sqrt{1-r^2} \cos k_n \right) 
\end{equation}
with $\lambda_n$ of Eq.~\eqref{eq:discrete_eigenvalues}.
The stationary-state solution of Eq.~\eqref{eq:mode_eom} for $n>0$ is
\begin{equation} \label{eq:mode_stat_state}
  \chi_{\alpha n}(t)  = e^{-t/\tau_n} \int_{-\infty}^{t} e^{t'/\tau_n}\left( \dot \gamma \delta_{\alpha x}\chi_{yn} (t') + \frac{1}{\zeta}  \Tilde\Gamma_{\alpha n}(t') \right) dt' , 
\end{equation}
where the relaxation times  $\tau_n= \zeta/\xi_n$ are introduced.  For $n =0$ follows 
\begin{equation}
  \chi_{\alpha 0}(t)  = \chi_{\alpha 0}(0) + \int_{0}^{t} \left( \dot \gamma \delta_{\alpha x}\chi_{y0} (t') + \frac{1}{\zeta}  \Tilde\Gamma_{\alpha 0}(t') \right) dt' .
\end{equation}

\subsection{Correlation Functions of Mode Amplitudes}  \label{sec:mode_correlations}

The correlation functions in the gradient and vorticity directions ($\alpha \in \{y,z\}$) do not explicitly depend on the shear rate. Hence, Eq.~\eqref{eq:mode_stat_state} yields~\cite{phil:22.1}
\begin{equation}
 \lla \chi_{n \alpha}(t)\chi_{m \alpha}(t') \rra  = \frac{2k_BT}{\xi_n + \xi_m} \bm b_n^{\dagger}  \cdot  \bm b_m^{\dagger} e^{-|t-t'|/\tau_x} ,   
\end{equation}
where $\tau_x=\tau_n$ for $t >t'$ and  $\tau_x=\tau_m$ for $t < t'$. Correlation functions, including the mode $n=0$, are given by 
\begin{align} \label{Eq:mode_corr}
\lla \chi_{\alpha n}(t) \chi_{\alpha 0}(t') \rra = & \ \frac{2k_BT}{\xi_n} \: \bm{b}_n^{\dagger} \cdot \bm{b}_0^{\dagger}
		\begin{dcases}
			e^{-(t-t')/\tau_n} , & t>t' 
			\\
			1 , & t \le t'
		\end{dcases}, 
		\\
	\lla \chi_{\alpha 0}(t) \chi_{\alpha 0}(t') \rra = & \ \frac{2k_BT}{\zeta} \: \bm{b}_0^{\dagger} \cdot \bm{b}_0^{\dagger} \: t' + \chi_{\alpha 0}^2(0), \ \ \  t \ge  t' \ge 0 .	
\end{align}
Correlation functions that include the flow direction are as follows: 
\begin{widetext}
\begin{align} \nonumber 
 \lla \chi_{x n}(t)\chi_{y m}(t') \rra  = & \ \dot \gamma \frac{2k_BT \zeta}{(\xi_n + \xi_m)^2} \bm b_n^{\dagger}  \cdot  \bm b_m^{\dagger}  \left[1 + \left(\frac{1}{\tau_n} + \frac{1}{\tau_m} \right)(t-t') \Theta(t-t') \right] e^{-|t-t'|/\tau_x},   \\[5pt]
  \lla \chi_{x n}(t)\chi_{x m}(t') \rra  =  & \ \lla \chi_{y n}(t)\chi_{y m}(t') \rra + {\dot \gamma^2} \frac{2k_BT \zeta^2}{(\xi_n + \xi_m)^3} \bm b_n^{\dagger}  \cdot  \bm b_m^{\dagger}  \left[2 + \left(\frac{1}{\tau_n} + \frac{1}{\tau_m} \right)(t-t')  \right] e^{-|t-t'|/\tau_x} .
\end{align}
\end{widetext}
As before, $\tau_x=\tau_n$ for $t >t'$ and  $\tau_x=\tau_m$ for $t < t'$. $\Theta(t)$ is the Heaviside step function.  

The mode-amplitude correlation function in the steady state, $\langle \bm \chi_n \cdot \bm \chi_m \rangle$, is then 
\begin{equation} \label{eq:corr_st_st}
\lla \bm \chi_n \cdot \bm \chi_m \rra =   \left[\frac{6 k_BT}{\xi_n + \xi_m} + \dot \gamma^2 \frac{4 k_BT \zeta^2}{(\xi_n + \xi_m)^3}\right] \bm b_n^{\dagger}  \cdot  \bm b_m^{\dagger} .
\end{equation}

\section{Stretching Coefficient and Relaxation Times}

\subsection{Stretching Coefficient}

A finite contour length is a fundamental property of a polymer and determines its conformational and
dynamical characteristics. \cite{wink:24,eise:16} The role of $\mu$ becomes particularly significant in the presence of external forces, such as a linear shear flow, \cite{wink:10,wink:24} since such forces attempt to stretch the polymer.\cite{wink:03,perk:95,mark:95} Insertion of the eigenfunction expansion into Eq.~\eqref{eq:const}  yields   
\begin{align} \label{eq:constraint_mu} \nonumber 
    N l^2 & =  \sum_{n,m=1}^N \lla \bm \chi_n \cdot \bm \chi_m \rra \sum_{i=1}^{N} (b_n^i-b_n^{i-1})(b_m^i-b_m^{i-1}) \\ \nonumber
    & = \sum_{n,m=1}^N \left[\frac{6 k_BT}{\xi_n + \xi_m} + \dot \gamma^2 \frac{4 k_BT \zeta^2}{(\xi_n + \xi_m)^3}\right] \bm b_n^{\dagger}  \cdot  \bm b_m^{\dagger}   \\
    & \ \ \ \times \sum_{i=1}^{N} (b_n^i-b_n^{i-1})(b_m^i-b_m^{i-1}),
\end{align}
with the equal-time correlation function of Eq.~\eqref{eq:corr_st_st}. In the following, the Weissenberg number 
\begin{equation} \label{eq:def_wi}
    Wi = \dot \gamma \tau_R  
\end{equation}
is used to characterize the shear-rate dependence, where $\tau_R = \zeta N^2 l^2/(3 \pi^2 k_BT)$ is the Rouse relaxation time.\cite{doi:86} 

\begin{figure}[t]
\includegraphics[width=\columnwidth]{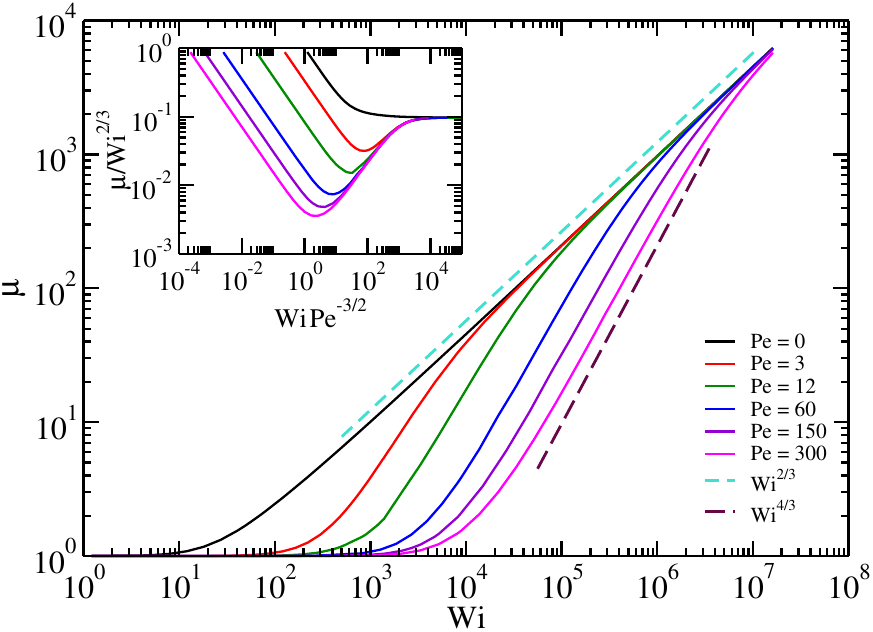}
	\caption{Stretching coefficient $\mu$ as function of the Weissenberg number $Wi$ (Eq.~\eqref{eq:def_wi}) for various P\'eclet numbers $Pe$ (legend) and the number of beads $N=200$. Inset: scaled stretching coefficient, illustrating the universal crossover from the activity-dominated to the passive shear-rate regime.}
	\label{fig:stretch}
\end{figure}

Figure~\ref{fig:stretch} presents the numerical solution of Eq.~\eqref{eq:constraint_mu}. In general, the stretching coefficient increases with increasing shear rate, indicating increasing bond-stretching forces, where the threshold depends on the P\'eclet number. In the passive limit $Pe = 0$, $\mu$ is given by $\mu=Wi^{2/3}/(6N)^{1/3}$ and  it increases with the power-law exponent $2/3$.\cite{wink:10} As activity increases, a different power-law-type regime appears for $Pe > 1$, where the stretching coefficient approximately obeys the relation $\mu \sim Wi^{4/3}$. The crossover from the shear-free, passive polymer value $\mu =1$ to this power-law regime shifts to large Weissenberg numbers with increasing shear rate. This indicates that the Rouse time is not the relevant relaxation time characterizing the onset of shear effects on the polymer. Since such effects appear for shear rates $\dot \gamma \tau_s \gtrsim 1$,  where $\tau_s$ is the relevant characteristic relaxation time, the shift of the curves to large $Wi$ indicates that $\tau_s$ decreases with increasing activity. As many modes contribute to the evaluation of Eq.~\eqref{eq:constraint_mu}, we cannot easily identify a single dominant relaxation time. However, the sums over modes in the shear-rate-dependent term of Eq.~\eqref{eq:constraint_mu} seem to become smaller with increasing P\'eclet number, and larger Weissenberg numbers are required for a shear response. 

At large Weissenberg numbers, the stretching coefficient $\mu$ approaches the value of the passive polymer. This suggests that shear dominates over activity. The crossover from the activity-dominated to the passive regime occurs for $r =Pe/(6 \mu) \ll 1$ and the eigenfunctions and eigenvalues assume their passive form. By setting $r \sim Pe/\mu = const. \ll 1 $ and assuming $\mu \sim Wi^{2/3}$, we find the crossover condition
\begin{equation}
Wi Pe^{-3/2} \gg 1 . 
\end{equation}
The inset in Fig.~\ref{fig:stretch} confirms this scaling prediction. 

In the absence of shear, $\mu =1$ independent of the P\'eclet number, and the polymer conformations are activity independent. The increase of $\mu$ with increasing $Wi$ is related with a decrease of $r$, until the passive limit with $r=0$ is reached. An estimation of the $Pe$-dependent shift of the various curves in Fig.~\ref{fig:stretch} can be obtained by setting $r=const.=1$. Insertion of $\mu \sim Wi^{4/3}$, yields then 
\begin{equation}
  Wi Pe^{-3/4} = const. 
\end{equation}
The scaling of $Wi$ confirms this prediction (cf. Sec.~\ref{sec:conf_gyrat}). Consequently, the relaxation time $\tau_s$, which characterizes the shear-induced polymer deformation, obeys the relation $\tau_s \sim Pe^{-3/4}$. 

\REV{Aside from the P\'eclet and the Weissenberg number, the stretching coefficient depends on the polymer length $N$. Considering various polymer lengths, we primarily observe a shift of the $\mu$ curves toward larger Weissenberg numbers with increasing $N$, a phenomenon also observed for passive polymers.\cite{wink:10} Similar to the dependence on $Pe$, the transition to passive polymer behavior occurs with increasing $Wi$ for shorter polymers at smaller $\mu$ values. An asymptotic behavior is only achieved for long polymers, $N \gtrsim 10^2$. Numerically, we find an overlap of the curves for the different $N$ by scaling the Weissenberg number with $N^{-4/3}$. Hence, the stretching coefficient approximately follows the relationship $\mu \sim (Wi/[Pe^{3/4} N^{4/3}])^{4/3}$ for $\mu \gg 1$ and $N \gg 1$. 
}

\subsection{Relaxation Times}

The relaxation times of the APLP follow from the eigenvalues in Eq.~\eqref{eq:eigenvalues}. Since $r = Pe/(6 \mu)$ can be greater than one for large $Pe$ and small $\mu$, the eigenvalues can be complex. In the regime $r<1$, $\zeta_n \in \mathbb{R}$, and the relaxation times are given by 
\begin{equation} \label{Eq:passive}
    \tau_n = \frac{\zeta}{\xi_n} = \frac{\pi^2 \tau_R}{2N^2 \mu \left[ 1 - \sqrt{1 - r^2} \, \cos k_n \right]} .
\end{equation}
For $r > 1$, $\xi_n$ becomes complex. With the definition 
\begin{equation}
\xi_n = \frac{\zeta}{\tau} -  \imath \zeta \omega_n ,    
\end{equation}
we obtain a single mode-independent relaxation time $\tau$ and mode-dependent frequencies $\omega_m$, which read 
\begin{equation}
    \tau = \frac{\pi^2}{2N^2 \mu} \tau_R, \quad
\omega_n = \frac{2N^2 \mu}{\pi^2 \tau_R} \sqrt{r^2 - 1} \, \cos k_n .
\label{Eq:active}
\end{equation}

\begin{figure}[t]
\includegraphics[width=\columnwidth]{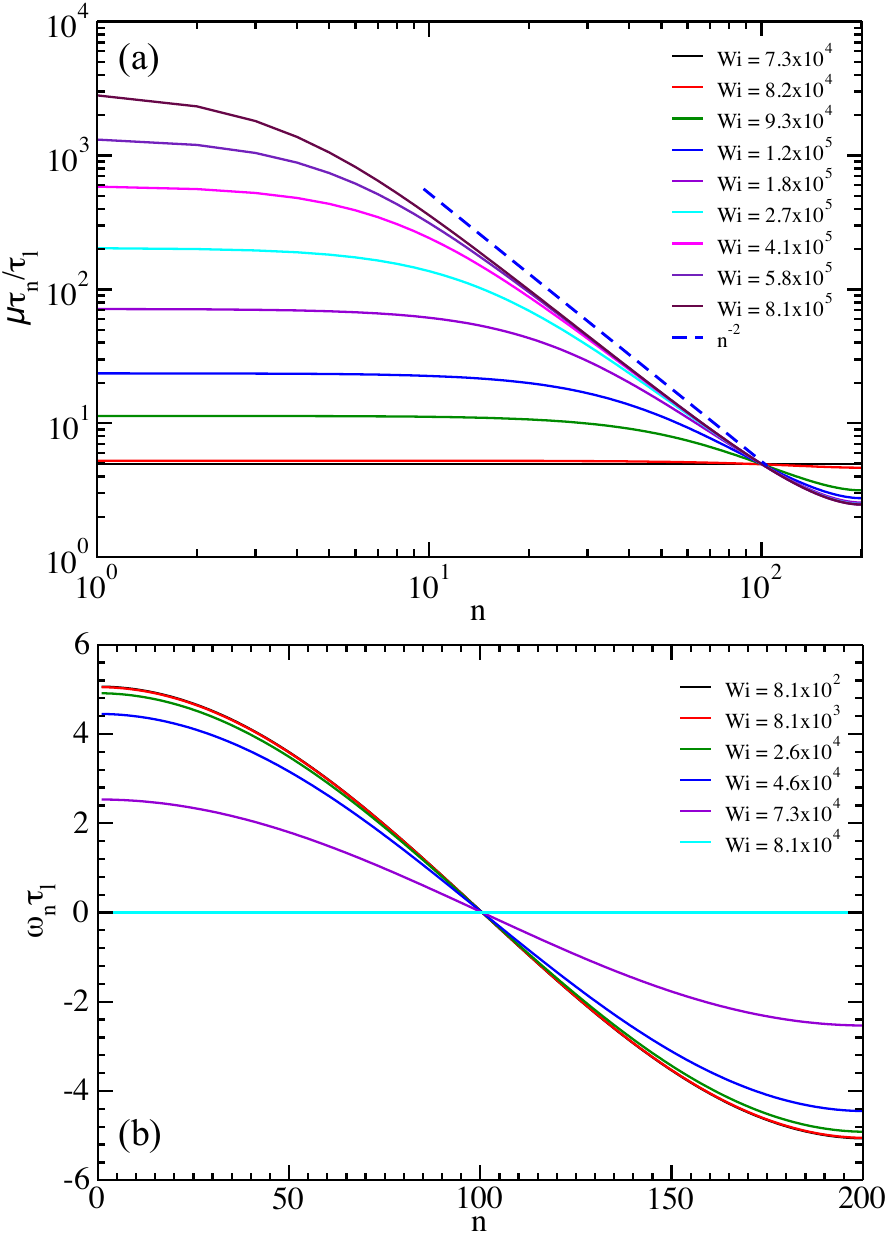}
	\caption{(a) Relaxation times $\mu \tau_n/\tau_l$ and \REV{(b) frequencies $\omega_n \tau_l$ as a function of the mode number $n$ for the P\'eclet number $Pe =150$ and various Weissenberg numbers $Wi$ (legend). The relaxation time $\tau_l$ is defined as $\tau_R/N^2$. The dashed line in (a) indicates the power law $1/n^2$.  }}
	\label{Fig:tau_n}
\end{figure}

Figure~\ref{Fig:tau_n}(a) shows relaxation times as a function of the mode number and various Weissenberg numbers. Since we are representing $\mu \tau_n$, the figure reflects the dependence of the relaxation times on $r=Pe/(6 \mu)$. The increasing times $\mu \tau_n$ with increasing $Wi$, reflect the crossover from the activity-dominated polymer dynamics for small $Wi$ and $r>1$ to the passive limit for $WiPe^{-3/2} \gg 1$ and $r \to 0$. In the latter case,  the relaxation times decay as  $1/n^2$ over a range of mode numbers. This range increases with increasing polymer length.   
As long as $r >1$, which is the case for $Wi \lesssim 9$ in Fig.~\ref{Fig:tau_n}(a), the product $\mu \tau_n = \mu \tau$ (Eq.~\eqref{Eq:active}) is independent of activity and shear.  The reverse applies to the frequencies (Fig.~\ref{Fig:tau_n}(b)); $\omega_n$ exhibits a Weissenberg-number dependence for $r>1$ and becomes zero for $r<1$. As long as $r>1$,  the magnitude of the frequencies, $|\omega_n|$, decreases with increasing shear rate, and for $r \gg 1$,  $\omega_n \sim (Pe^2 \cos k_n)/\mu$ (Eq.~\eqref{Eq:active}).    


\section{End-to-End Vector Correlation Function}

A P\'eclet-number-dependent characteristic relaxation time of the APLP can be obtained from the end-to-end vector correlation function $\langle \bm{R}_e(t) \cdot \bm{R}_e(0) \rangle$, with the end-to-end vector $\bm R_e = \bm r_N - \bm r_0$. In terms of the eigenfunction expansion, the correlation function in the absence of shear reads 
\begin{widetext}
    \begin{equation}  \label{eq:end_corr}
    	 C_e(t) = \frac{\lla \bm{R}_e(t) \cdot \bm{R}_e(0) \rra}{\lla \bm R_e^2 \rra}  =  \frac{6 k_BT}{Nl^2}  \sum_{n,m=1}^{N}  \frac{\bm{b}_n^{\dagger} \cdot \bm{b}_m^{\dagger}}{\xi_n+\xi_m} \left[b_n^{(N)} - b_n^{(0)}\right] \left[ b_m^{(N)} - b_m^{(0)} \right] e^{- \xi_n t/  \zeta} ,
    \end{equation}
\end{widetext}
with the equilibrium mean-square end-to-end distance $\langle \bm R_e^2 \rangle = N l^2$.\cite{phil:22.1}

\begin{figure}[t]
\includegraphics[width=\columnwidth]{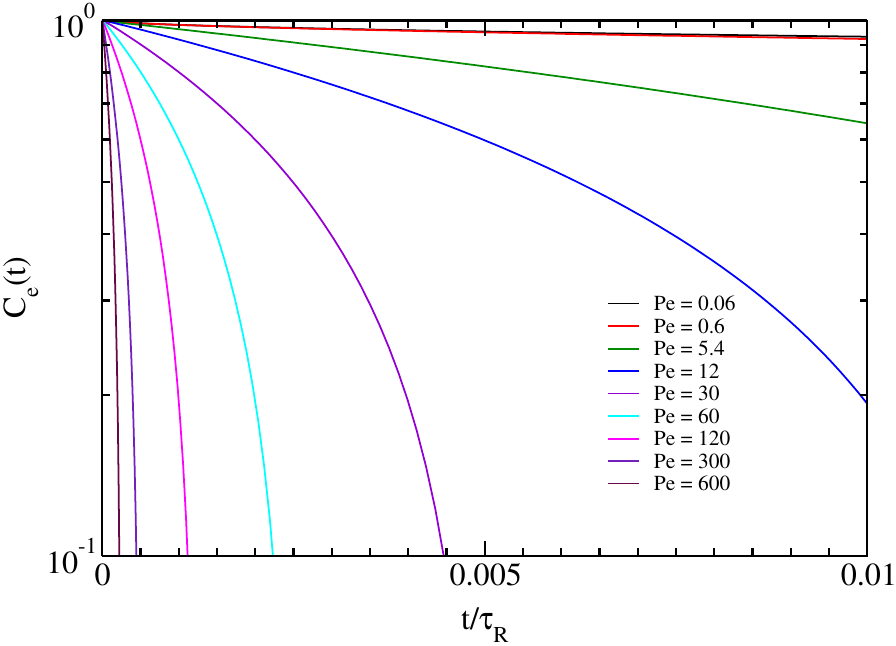}
	\caption{Normalized polymer end-to-end vector correlation function $C_e(t)$ (Eq.~\eqref{eq:end_corr}) as a function of the scaled time $t/\tau_R$ for the polymer length $N=200$ and various P\'eclet numbers (legend). }
	\label{Fig:corr}
\end{figure}

Figure~\ref{Fig:corr} depicts the correlation function $C_e(t)$ for various P\'eclet numbers. The $Pe$ range encompasses $r$ values that are both smaller and larger than one. The correlation function decays in a nonexponential manner for a wide spectrum of $Pe$ values. Hence, typically, many modes contribute to the relaxation. A nonexponential decay has been predicted analytically in Ref.~\onlinecite{phil:22.1}, and has been observed in various simulations. \cite{faze:23,teje:24,pand:25,panda2025folding} As long as $r <1$, $C_e$ decays exponentially at long times, with the largest relaxation $\tau_1$. \cite{phil:22.1} For $r=Pe/(6\mu) >1$, relaxation is not determined by the single relaxation time $\tau$, but rather by the oscillation frequencies $\omega_n$. 

\begin{figure}[t]
\includegraphics[width=\columnwidth]{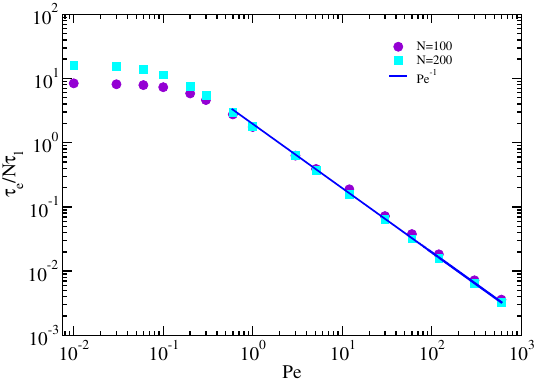}
	\caption{Scaled characteristic relaxation time $\tau_e$ of APLPs in absence of shear as a function of $Pe$ for the polymer lengths  $N=100$ and $N=200$. The solid line (blue) indicates the power $\tau_e \sim Pe^{-1}$. The relaxation time $\tau_l=\tau_R/N^2 = \gamma l^2/(3k_BT)$. }
	\label{Fig:time}
\end{figure}

We define a characteristic relaxation time $\tau_e$ by the time, where the correlation function has decayed to $1/e$, i.e., $C_e(\tau_e) =1/e$. The attempt to fit a stretched exponential function was unsuccessful. The obtained $\tau_e$ values are presented in Fig.~\ref{Fig:time}. For small $Pe< 1$, $\tau_e$ is independent of activity and is given by the Rouse relaxation time. In the regime $Pe > 1$, the relaxation time follows the power law $\tau_e \sim Pe^{-1}$, as has already been observed in Ref.~\onlinecite{pand:25}. A similar dependence has been obtained in Refs.~\onlinecite{faze:23,teje:24}, applying different criteria to extract a relaxation time. Moreover, we confirm the polymer-length dependence $\tau_e \sim N$.\cite{pand:25,panda2025folding,faze:23,teje:24} 

As mentioned above, it is difficult to obtain an analytical expression for the relaxation time $\tau_e$, since in general many modes contribute to the decay of the correlation function $C_e$. Alternatively, a characteristic relaxation time $\tau_p$ can be defined as the time at which the polymer dynamics reaches a stationary state. A suitable quantity is the average bead mean-square displacement (MSD) in the polymer center-of-mass reference frame. The stationary state is reached when this displacement reaches a time-independent value, which, for the current model, is $\langle \bm R_e^2 \rangle/3$. For longer times, the total polymer MSD is given by its center-of-mass MSD, which increases linearly with time and the center-of-mass diffusion coefficient (cf. Ref.~\onlinecite{phil:22.1}). Equating the MSD of the polymer center-of-mass and the plateau value of the bead MSD in the center-of-mass reference frame yields 
\begin{equation} \label{eq:time_p}
    \tau_p = \frac{\gamma l^2 N}{3 k_B T Pe}.
\end{equation}
This estimate reflects the relaxation-time dependencies on $Pe$ and $N$ observed in simulations, \cite{faze:23,teje:24,pand:25,panda2025folding} and confirms that the time $\tau_e$ is suitable for characterizing the overall relaxation of the active polymer.

\section{Conformational Properties}

\subsection{Radius of Gyration } \label{sec:conf_gyrat}

The polymer conformations are characterized by the radius-of-gyration tensor \cite{huan:10,wink:10,pand:25}
\begin{equation}
    G_{\alpha \beta} = \frac{1}{N+1}\sum_{i=0}^N \lla \Delta r_{\alpha i} \Delta r_{\beta i} \rra ,
\end{equation}
where $\langle \Delta \bm r_{i} \rangle$ is the position of bead $i$ with respect to the polymer's center of mass.  Insertion of the expansion Eq.~\eqref{eq:exp_pos} yields
\begin{equation}
    G_{\alpha \beta} =  \sum_{n,m=0}^N   \lla \chi_{\alpha n} \chi_{\beta m} \rra   A_{nm} ,
\end{equation}
with the abbreviation 
\begin{equation}
    A_{nm} = \frac{1}{N+1} \left[\bm b_n \cdot \bm b_m - \frac{1}{N+1}
    \sum_{i,j =0}^N b_n^i b_m^j \right] .
\end{equation} 
Note that $A_{nm}=0$ for $n=0 \lor m=0$. 
Explicitly, the components along the flow, $G_{xx}$, and the gradient, $G_{yy}$, direction read
\begin{align}
G_{xx} =  & \   G_{yy} + {\dot \gamma^2} \sum_{n,m=1}^N  \frac{4k_BT \zeta^2}{(\xi_n + \xi_m)^3} \ \bm b_n^{\dagger}  \cdot  \bm b_m^{\dagger}  A_{nm}  , \\
G_{yy} = & \  \sum_{n,m=1}^N \frac{2k_BT}{\xi_n + \xi_m} \ \bm b_n^{\dagger}  \cdot  \bm b_m^{\dagger}  A_{nm} ,
\end{align}
with the mode-amplitude correlation functions of Sec.~\ref{sec:mode_correlations}.

\begin{figure}[t]
\includegraphics[width=\columnwidth]{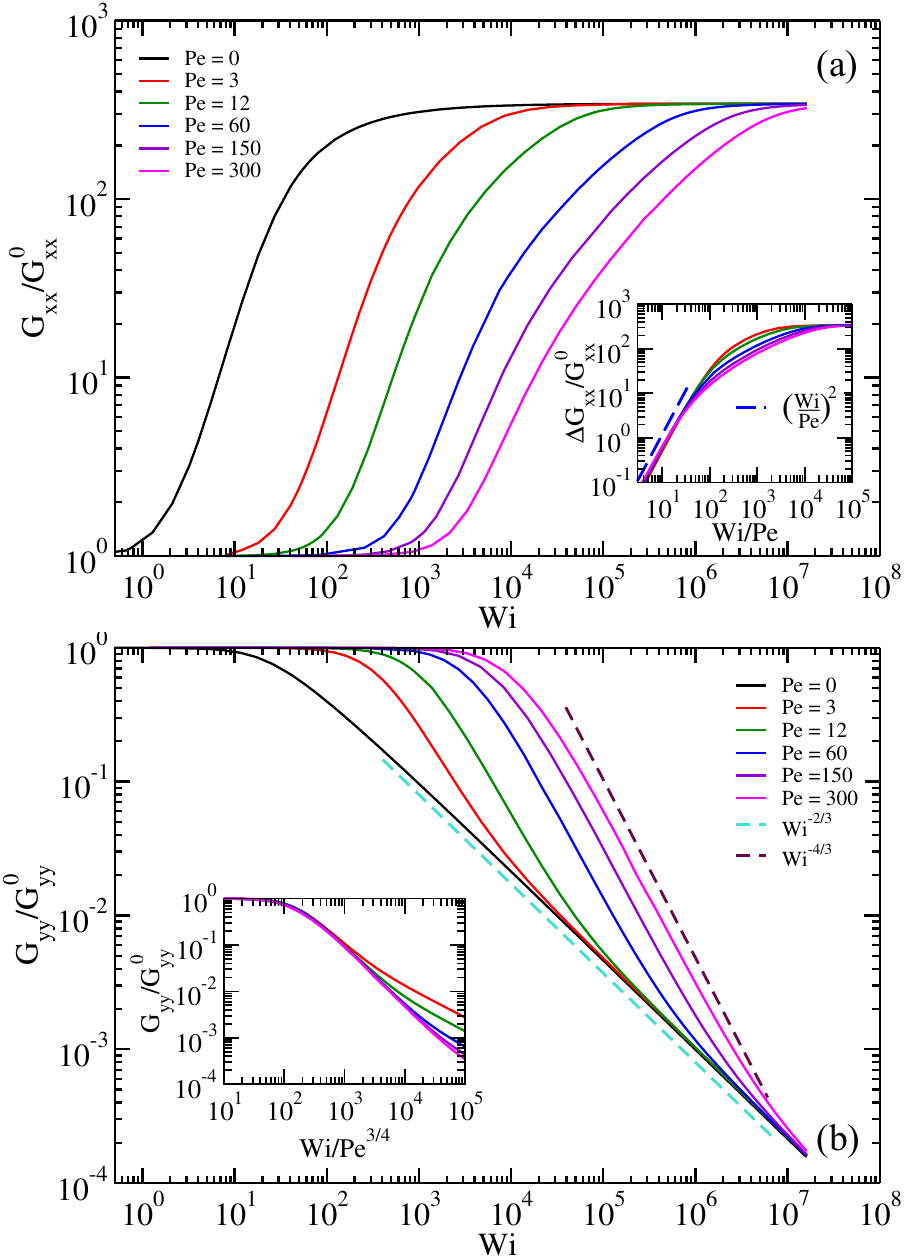}
	\caption{Normalized radius-of-gyration tensor (a) $G_{xx}/G_{xx}^0$ along the flow direction and (b)  $G_{yy}/G_{yy}^0$ along the gradient direction as a function of the Weissenberg number $Wi$ for various $Pe$ (legend). The components $G_{xx}^0$ and $G_{yy}^0$ are the zero-shear values. The dashed lines in (b) indicate power laws with the exponents $2/3$ (passive limit, $Pe=0$) and $4/3$ ($Pe >0$), respectively. The number of beads is $N=200$. Insets: (a) $\Delta G_{xx}/G_{xx}^0 = G_{xx}/ G_{xx}^0-1$ as a function of the scaled Weissenberg number $Wi/Pe$. The dashed line shows the quadratic increase of $\Delta G_{xx}$ with increasing shear rate.  (b) $G_{yy}$ as a function of the scaled Weissenberg number $Wi/Pe^{3/4}$.}
	\label{Fig:gxx}
\end{figure}

Figure~\ref{Fig:gxx}(a) presents the radius-of-gyration tensor components along the flow and gradient directions as a function of the Weissenberg number and various P\'eclet numbers. The component $G_{xx}$ (Fig.~\ref{Fig:gxx}(a)) along the flow direction increases relative to the equilibrium (zero-shear) value $G_{xx}^0$ with increasing flow strength, and assumes a $Pe$-independent plateau value, when $\mu$ assumes the passive limit. Consequently, the plateau value is equal to the value of a passive Gaussian polymer under shear flow.\cite{wink:10} Activity causes a shift of the curves toward larger Weissenberg numbers. For small Weissenberg numbers, as long as $\mu \approx 1$, $\Delta G_{xx} =G_{xx} - G_{xx}^0$ increases quadratically with $Wi$, as can be deduced from Eq.~\eqref{eq:corr_st_st}, and is illustrated in the inset of Fig.~\ref{Fig:gxx}(a). The polymer remains undeformed but gradually aligns along the direction of flow. The shift toward larger Weissenberg numbers with increasing $Pe$ is caused by the respective decrease of the relaxation time $\tau_p$ (Eq.~\eqref{eq:time_p}) (or $\tau_e$). This is confirmed in the inset of Fig.~\ref{Fig:gxx}(a).   

Polymer stretching along the flow direction is invariably accompanied by shrinkage in the transverse directions, as depicted in Fig.~\ref{Fig:gxx}(b). It is noteworthy that such a stretching of the APLP occurs at higher Weissenberg numbers than the increase in $G_{xx}$, as can be inferred from a comparison of Figs.~\ref{Fig:gxx}(a) and (b). Since  $G_{yy}$ is proportional to $1/\mu$, and does not explicitly depend on the shear rate, it closely resembles the $Wi$ and $Pe$ dependence of the stretching coefficient. Hence, $G_{yy}$ obeys the scaling relation $Wi Pe^{-3/4}$ with respect to $Wi$ and $Pe$, as is confirmed in the inset in Fig.~\ref{Fig:gxx}(b).
In the passive limit, $\mathrm{Pe} \to 0$, the polymer shrinkage follows the scaling law $G_{yy}/G_{yy}^0 \sim Wi^{-2/3}$ in the asymptotic shear regime $Wi \to \infty$.\cite{wink:10,huan:10}. In the presence of activity, as for the stretching coefficient, the scaling regime $G_{yy}/G_{yy}^0 \sim Wi^{-4/3}$ appears, which crosses over to the passive polymer limit for $WiPe^{-3/2} \gg 1$.

The difference in the shift of the onset of the conformational changes along the flow and gradient directions suggests that the two quantities are determined by different relaxation processes. The growth of $G_{xx}$ occurs before a substantial deformation of the polymer emerges. The anisotropic polymer coil is aligned by shear flow, thereby, the polymer is fully relaxed. Hence, this process is determined by the longest relaxation $\tau_p$ (Eq.~\eqref{eq:time_p}). The deformation and shrinkage of the active polymer, which appears at larger Weissenberg numbers, is determined by various or many shorter relaxation processes. This leads to the distinct difference in the P\'eclet-number dependence of the effective Weissenberg numbers along and transverse to the flow direction.         

\section{Alignment} \label{sec:alignment}

Shear induces alignment of the polymer along the flow direction.\cite{smit:99,aust:99,wink:10,huan:10,pand:25} This alignment can be quantified by the angle $\chi$ between the principal eigenvector of the radius-of-gyration tensor with the largest eigenvalue and the flow direction. It can be expressed in terms of the components of the radius-of-gyration tensor as  follows: 
\begin{align} \nonumber
\tan(2\chi)  & \  =\frac{2G_{xy}}{G_{xx} - G_{yy}} \\ & \ = \frac{1}{\dot \gamma} \frac{\displaystyle \sum_{n,m=1}^{\infty} \bm b_n^{\dag} \cdot \bm b_m^{\dag} A_{nm} /(\xi_n+\xi_m)^2}{\displaystyle \sum_{n,m=1}^{\infty}  \zeta \bm b_n^{\dag} \cdot \bm b_m^{\dag} A_{nm} /(\xi_n+\xi_m)^3}.
\label{eq:alignment}
\end{align}

\begin{figure}[t]
\includegraphics[width=\columnwidth]{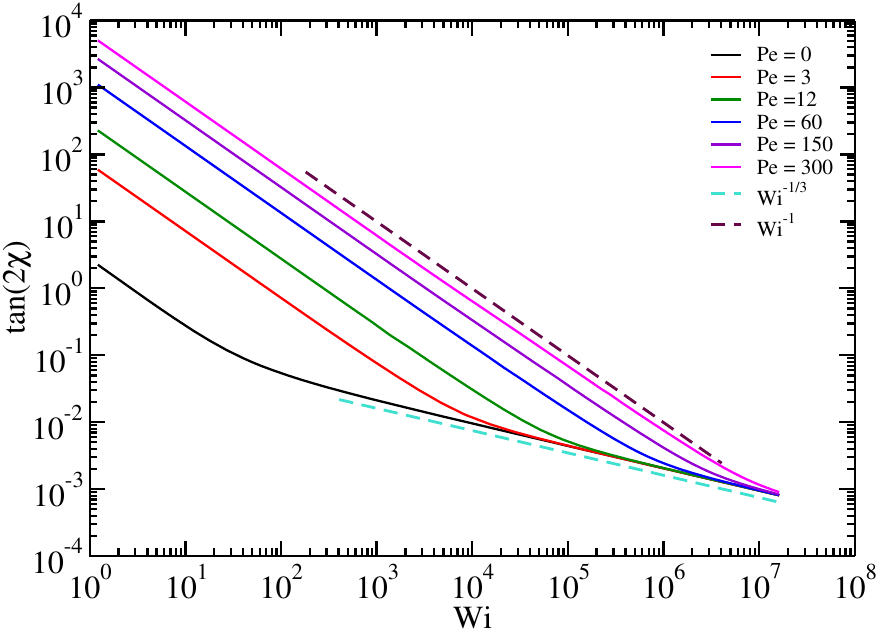}
	\caption{Shear-induced alignment of APLPs as a function of the Weissenberg number $Wi$ and various activity strengths $Pe$ (legend) for $N=200$.}
	\label{Fig:Align}
\end{figure}

Figure~\ref{Fig:Align} illustrates the polymer’s alignment as a function of the Weissenberg number and for various activities. The different curves show two distinct power-law regimes. As long as the polymer is unstretched, $\mu \gtrsim 1$, $\tan(2\chi)$ decreases as $1/Wi$ with increasing shear rate. This follows from Eq.~\eqref{eq:alignment}, since $\tan(2\chi) \sim 1/Wi$. The range of this power-law regime increases with  P\'eclet number, despite the shear-induced changes  in polymer conformations. As $G_{xx}$ grows with increasing Weissenberg number, $G_{xy}$ crosses over from the shear-rate dependence of the undeformed polymer, $G_{xy} \sim Wi$, to the asymptotic, passive polymer decay $G_{xy} \sim Wi^{1/3}$ with increasing shear rate. The combined effect yields the extended alignment regime with the $1/Wi$ decay. This is a consequence of the difference in the onset of the shear-induced conformational changes along the flow and the transverse direction---$G_{xy}$ shows the same scaling relation $WiPe^{-3/4}$ as the stretching coefficient and $G_{yy}$ for the onset of the shear effect.          
In the passive polymer limit $\tan(2\chi) \sim Wi^{-1/3}$ for $Wi \gg 1$.\cite{huan:12,wink:10} The simulations of APLPs has displayed similar characteristics.\cite{pand:25}

\section{Viscosity}

The rheological properties of polymers are tightly linked with their flow-induced conformational changes. \cite{bird:87,aust:99,teix:05,schr:05.1,huan:10,wink:10,mart:18.1,pand:23,wink:24,pand:25} Here, we focus on the dependence of the 
shear viscosity on the active force and the shear rate.

The polymer contribution to the viscosity is calculated via the virial expression for the stress tensor,\cite{bird:87,wink:10,pand:25} which contains the bond, $\bm F_i^b \sim \mu$, and active, $\bm F_i^a \sim f_a$, forces of the APLP (cf. Eqs.~\eqref{eq:eom_0} - \eqref{eq:eom_n}). Explicitly, the stress tensor, $\sigma_{xy}$, is given by  
\begin{equation}
    \sigma_{xy} = - \frac{1}{V} \sum_{i=0}^{N}  \lla \left( F_{x i}^b + F_{x i}^a \right) r_{y i} \rra .
\end{equation}
Insertion of the forces and the eigenfunction expansion yields the polymer contribution to the shear viscosity $\eta_p= |\sigma_{xy}|/\dot \gamma$: 
\begin{equation}
    \eta_p = \frac{2 k_BT \zeta}{V} \sum_{n=1}^N \sum_{m=0}^N \frac{\xi_n}{(\xi_n + \xi_m)^2} (\bm b_n \cdot \bm b_m) (\bm b_n^\dag  \cdot \bm b_m^\dag) . 
\end{equation}

\begin{figure}[t]
\includegraphics[width=\columnwidth]{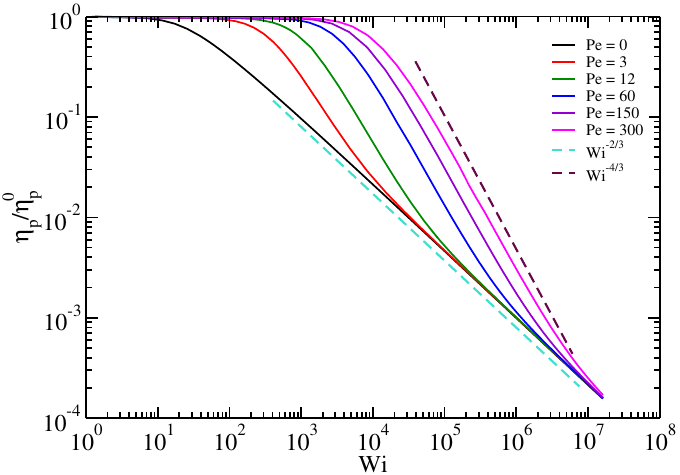}
	\caption{Normalized shear viscosity $\eta_p/\eta_p^0$ of APLPs as a function of the Weissenberg number $Wi$ for various activities $Pe$ at polymer length $N=200$. The dashed lines indicate the power-law behaviors in presence ($\eta_p \sim Wi^{-4/3}$, light blue) and absence ($\eta_p \sim Wi^{-2/3}$, black) of activity. Here, $\eta_p^0$ corresponds to the zero-shear viscosity. }
	\label{Fig:visc}
\end{figure}

Figure~\ref{Fig:visc} shows the shear-rate-dependent polymer viscosity, normalized by its activity-dependent zero-shear value $\eta_p^0$, as a function of the Weissenberg number. Our calculations reveal a strong activity dependence, with a significantly enhanced shear-thinning behavior compared to a passive polymer. The slope changes from the passive-polymer value\cite{wink:10} $\eta \sim Wi^{-2/3}$ to the activity-dominated value $\eta \sim Wi^{-4/3}$ for $Pe \gg 1$ over an activity-dependent range of shear rates. In the asymptotic limit $Wi Pe^{-3/2} \gg 1$, the viscosity assumes the passive value for any P\'eclet number. This has also been found in simulations.\cite{pand:25}      

Strikingly, the viscosity closely displays the shear-rate and activity dependence of the stretching coefficient $\mu$. Since $\xi_n \sim \mu$, $\eta_p \sim 1/\mu$ aside from an additional dependence of the $\xi_n$, $\bm b_n^\dag$, and $\bm b_n^{\dag}$ on $\mu$ via $r$. Correspondingly, the onset of the drop of $\eta_p$ with increasing shear rate scales as $WiPe^{-4/3}$. The activity and shear-rate dependence of $\eta_p/\eta_p^0$ and $G_{yy}/G_{yy}^0$ is very similar, because both quantities are predominantly determined by $\mu$. This is particularly evident for passive polymers under shear flow, where those quantities are essentially given by the longest polymer relaxation time.\cite{wink:10} Since the stretching coefficient reflects the forces along the polymer contour, the alignment and the shear viscosity are determined by these forces.


\section{Summary and Discussion}

We have analyzed the behavior of flexible APLPs in shear flow within an analytical framework. The active phantom polymer is modeled as a discrete, finite-extensible Gaussian chain, \cite{wink:94} driven by active bond forces.\cite{phil:22.1,wink:25} Excluded-volume interactions are neglected to ensure analytical tractability. The linear, non-Hermitian equations of motion are solved using a biorthonormal basis expansion, and the finite extensibility of the chain is incorporated through a constraint force characterized by the Lagrange multiplier (stretching coefficient) $\mu$.\cite{wink:10,eise:16} The APLPs exhibit qualitatively a similar behavior as passive polymers under shear flow, namely stretching and alignment along the flow direction, shrinkage transverse to this direction, and shear thinning.  However, activity implies major quantitative differences.  

In the absence of shear, the conformations of the phantom APLPs are independent of activity and the stretching coefficient $\mu \equiv 1$.\cite{phil:22.1,wink:25} However, under applied shear, the conformations show a pronounced P\'eclet-number dependence, which is reflected in an increase of the stretching coefficient with increasing Weissenberg number with the approximate power-law $\mu \sim Wi^{4/3}$ in a $Pe$-dependent shear-rate regime. Remarkably, for $Wi/Pe^{2/3} \gg 1$ shear dominates over activity and the stretching coefficient exhibits passive polymer behavior, where $\mu \sim Wi^{2/3}$. The stretching coefficient determines the polymer conformations. In particular, the polymer size transverse to the flow direction---measured by the radius-of-gyration tensor component---scales approximately as $G_{yy} \sim 1/\mu$. Similarly, the polymer contribution to the shear viscosity follows the same scaling relation, $\eta_p \sim 1/\mu$. This reflects a tight link between polymer shrinkage transverse to the flow direction and the shear viscosity, which also applies to passive polymers.\cite{wink:10} The analytical predictions are in agreement with coarse-grained simulation studies in the presence of excluded-volume interactions.\cite{pand:25} In particular, the same scaling exponents have been obtained, which suggests that excluded-volume interactions hardly affect the scaling exponents and polymer properties in three dimensions under shear. 

\REV{A priori, it is not evident that the current theoretical approach applies to flexible polymers with excluded-volume interactions in two dimensions, specifically at small shear rates. As has been demonstrated, polar activity has a severe impact on the polymer conformations and their dynamics in two dimensions.\cite{isel:15,kais:14} Yet, excluded-volume interactions become less important at higher shear rates, and we expect the current approach to also apply in this limit.}   


In addition, we predict the scaling relation $Wi/Pe^{3/4}$ for the onset of the shear-induced polymer shrinkage and shear-thinning behavior. Hence, the Rouse time $\tau_R$ of a passive polymer is not the relevant relaxation time governing the onset of these shear-induced quantities. Rather, multiple modes contribute to $\mu$. Yet, the onset of stretching along the flow direction, which scales as $Wi/Pe$, is governed by a different relaxation time. Our studies suggest that this stretching is controlled by the longest polymer relaxation time. We determined this relaxation time by considering the relaxation of the polymer end-to-end distance, $\tau_e$, and the time at which the mean-square displacement of the beads, $\tau_p$, reaches a plateau value in the absence of shear. Both quantities exhibit the polymer length and P\'eclet-number dependence $N/Pe$, dependencies that have been obtained in simulations before. \cite{pand:25,teje:24,faze:23}   

The linear active polymers reveal pronounced qualitative and quantitative differences to active Brownian polymers (ABPOs) as well as active polar ring polymers (APRPs) under shear flow. As far as active Brownian polymers (ABPOs) are concerned, APLPs exhibit activity effects in a shear-rate-dependent regime only, contrary to ABPOs. Their shear response extends up to infinitely high shear rates as far as shrinkage transverse to the flow direction, alignment, shear thinning, and tumbling are concerned.\cite{mart:18.1,pand:23} However, polar activity leads to a significantly stronger shrinkage and shear thinning.    

\REV{The fundamental influence of polarity and geometry on the properties of polymers under shear is reflected in the distinct differences between APRPs and APLPs. Although the Langevin equations of polar ring polymers are non-Hermitian as well, the periodic boundary conditions lead to rather different eigenfunctions and eigenvalues.\cite{wink:24} This is related to the fact that the total active force on a ring is zero. As a consequence, the stationary-state properties of semiflexible phantom APRPs are independent of activity, and their conformational and rheological properties are equal to those of passive polymers under shear flow.\cite{wink:25} Yet, their dynamics under shear depends on activity.\cite{wink:25,kuma:23.1} In particular, a tank-treading-like motion emerges at long relaxation times and high activities, specifically for stiff rings.\cite{wink:24} For flexible APRPs, relaxation processes dominate over activity-induced tank treading, and the rings exhibit tumbling motion. Shear strongly affects the crossover from a tank-treading to a relaxation-dominated dynamics, and the ring polymers exhibit tumbling motion at high shear rates.\cite{wink:24}}

There are various ways to extend the current study. The article focuses on flexible polar polymers. Actin filaments or microtubules are rather stiff, hence, a description within a semiflexible polymer approach is more suitable. Adopting the Gaussian semiflexible polymer model\cite{harn:95,wink:10} still leads to linear equations of motion. Yet, the analysis is more laborious. 

Long-range hydrodynamic interactions are omitted in the current analytical study. \REV{Simulations of externally driven semiflexible polymers with tangential forces reveal a strong influence of fluid-mediated interactions on the polymers' conformational and dynamical properties.\cite{stei:24} Specifically, semiflexible polymers exhibit a pronounced shrinkage. Alternatively, an active polymer can be composed of self-propelled beads. Studies of active Brownian polymers with externally driven\cite{mart:20} or self-propelled beads \cite{mart:19,wink:10} display qualitative differences in their conformations and dynamics. Analytically, fluid-mediated interactions may be accounted for by the preaveraging approximation of the hydrodynamic tensor.\cite{doi:86,wink:10,mart:19} This yields an additional mode-number dependence of the relaxation times. Since many modes generally contribute to the conformational properties of the APLPs under consideration, the influence of hydrodynamic interactions on their properties under shear is not immediately obvious.  Theoretical and simulation studies that account for hydrodynamic interactions may provide a more complete description of stiff filaments, microscopic robotic chains,\cite{caprini2024spontaneous}, or prototype filaments.
}

\begin{acknowledgments}

SPS and AP acknowledge funding support from the DST-SERB Grant No. CRG/2020/000661, and computational time at IISER Bhopal and the Param Himalaya NSM facility.
\end{acknowledgments} 


\section*{Conflict of Interest}

The authors declare no conflicts of interest.






\appendix

\section{Eigenvalue Problem of the APLP Model} \label{app:eigenvalues}

The Euclidean coordinates of the Eqs.~\eqref{eq:eom_0}--\eqref{eq:eom_n} are independent. Hence, only the $x$ direction is considered to find a solution of the eigenvalue problem.\cite{phil:22.1} Introducing the vector $\bm x = (x_0, \ldots, x_N)^T$, the equations of motion without shear force and thermal noise can be written as
\begin{align} \label{eq:eom}
    \dot {\bm x} =  - \frac{3k_BT \mu}{\zeta l^2}  \mathrm{\bf M}  \bm x , 
\end{align}    
with the tridiagonal matrix
\begin{widetext}
\begin{align} \label{eq:matrix}
    \mathrm{\bf M}        = 
    \begin{pmatrix}
        (1+r)&  -(1+r) & 0  & 0  &  0 &\cdots & 0 \\
        -(1-r)  & 2 & - (1+r)& 0&  0 & \cdots & 0 \\
        0 &  -(1-r) & 2 &  - (1+r) & 0 &\cdots &  0 \\
        \vdots  &  \vdots  & \begin{rotate}{382}$\ddots $
        \end{rotate}
        & 
        \begin{rotate}{382}$\ddots $
        \end{rotate}
        & \begin{rotate}{382}$\ddots $
        \end{rotate}
        & \vdots & \vdots \\
        0  & \cdots & 0 & -(1-r) & 2 &  - (1+r) & 0  \\
        0 &  \cdots & 0 & 0 &  -(1-r) & 2 &  - (1+r)  \\
        0  &   \cdots& 0 & 0  & 0& -(1-r)  & (1-r)  
    \end{pmatrix} ,
\end{align}    
\end{widetext}
and the abbreviations $r =  f_a l^2/(6 k_BT \mu)=Pe/(6 \mu)$ with $Pe=f_a l^2/(k_BT)$. The eigenvalues $\lambda_n$ and eigenvectors $\bm b_n = (b_n^{0}, \ldots, b_n^{N})^T$ of
the eigenvalue equation, 
\begin{equation}
    \mathrm{\bf M} \bm b_n =  \lambda_n \bm b_n ,
\end{equation}
follow via Laplace expansion.\cite{phil:22.1} The eigenvectors $\bm b_n^{\dagger}$ of the transposed matrix $\mathrm{\bf M}^{T}$,  with the same eigenvalue $\lambda_n$, are required for a complete basis set. 
Explicitly, the eigenvalues $\lambda_n$ are given by ($n=1,..., N$)
\begin{align} \label{eq:discrete_eigenvalues}
	\lambda_n = 2 \left( 1 - \sqrt{1-r^2} \cos k_n \right) \ ,
	\hspace{0.5cm} 
	\lambda_0 = 0,
\end{align}	
with the wave numbers $k_n = n \pi / (N+1)$. 

\end{document}